\documentclass{article}
\usepackage{graphicx} % Required for inserting images
\usepackage[a4paper, total={6.25in, 9in}]{geometry}
\usepackage{multirow}
\usepackage{amsmath}
\usepackage{float}
\usepackage{subcaption}

\title{The Elastic Analysis Facility’s (EAF’s) Contribution to the Future of Analysis at Multi-Experiment Institutions and Future Colliders \\[10pt]}

\author{Elise Chavez$^1$, Maria Acosta-Flechas$^2$, Christophe Bonnaud$^2$, \\ Burt Holzman$^2$, and Tulika Bose$^1$}
\date{%
    \small{$^1$University of Wisconsin-Madison} \\%
    \small{$^2$Fermilab}\\[2ex]%
    \today}

\begin{document}

\maketitle

\begin{abstract}
The Elastic Analysis Facility (EAF) hosted at Fermi National Accelerator Laboratory (Fermilab) is a platform being developed with the goal of providing a fast and efficient facility for physics analysis. As high-energy physics moves towards collecting larger datasets, such as those from the High-Luminosity LHC, the EAF strives to provide a powerful and adaptable framework for future colliders and multi-experiment institutions. Currently, the EAF supports several experiments including CMS, NOvA, and DUNE as well as serving accelerator physicists and beam line operations through integrated software and secure connections to Fermilab's computing resources. In addition, the EAF was designed with a user-friendly interface, intended to be more intuitive for emerging generations of physicists, that is still accessible for established styles of analysis. The EAF can also achieve better analysis efficiency due to the modernization of software and tools that can better utilize Fermilab's computing power. Furthermore, its design incorporates industry standards whenever possible, enhancing its sustainability and making it a possible template for other national or international laboratories and research facilities.  Overall, the EAF is a forward-looking solution that will meet the evolving needs of particle physics, ensuring readiness for future colliders and multi-experiment research institutions.
\end{abstract}

\newpage

\section{Introduction to the Elastic Analysis Facility}

\subsection{What is an Analysis Facility?}
High-energy particle (HEP) physics analysis is going through fundamental changes as the field moves to collect larger amounts of data from the Large Hadron Collider (LHC) with the High Luminosity LHC (HL-LHC) upgrade. Current analysis methods are relatively inefficient given the sheer volume of data and the complexities of developing an analysis. These challenges have prompted research into new analysis methods and software tools. Some of the inefficiencies originate from time lost in analysis development, data access, and data processing. Addressing these needs requires faster analysis development and improved tooling. Analysis facilities offer possible solutions for these issues and more.

The HEP community is still refining the definition and scope of an analysis facility; however, for the purposes of this paper, it can be loosely thought of as a suite of tools aimed at making a coherent analysis ecosystem, using software like JupyterHub and systems such as Kubernetes. These facilities enable users to utilize distributed computing, Python libraries and Pythonic analysis frameworks to process data faster. The ultimate goal of an analysis facility is to provide the user the ability to design, develop and run their analysis via distributed computing, specifically high throughput computing (HTC). It is a one-stop shop for a researcher that does not require them to be well equipped at software, tooling, or distributed computing. Analysis facilities can be configured in a variety of ways, depending on the established analysis pipelines and existing hardware. For example, some facilities (such as the U.S. CMS\footnotemark[1] MIT\footnotemark[2] facility) are built on non-Kubernetes platforms. More commonly, facilities are built on Kubernetes clusters with JupyterHub and other various software including SLURM, HTCondor, ServiceX, NVidia Triton, and many others. Examples of these facilities include the EAF and Coffea-Casa {\cite{coffea-casa}. A diagram of these types of facilities is given below in Figure \ref{eaf-af-diagram}.

\footnotetext[1]{CMS stands for Compact Muon Solenoid.}
\footnotetext[2]{MIT stands for Massachusets Institute of Technology.}

\begin{figure}[H]
    \centering
    \includegraphics[scale=0.3]{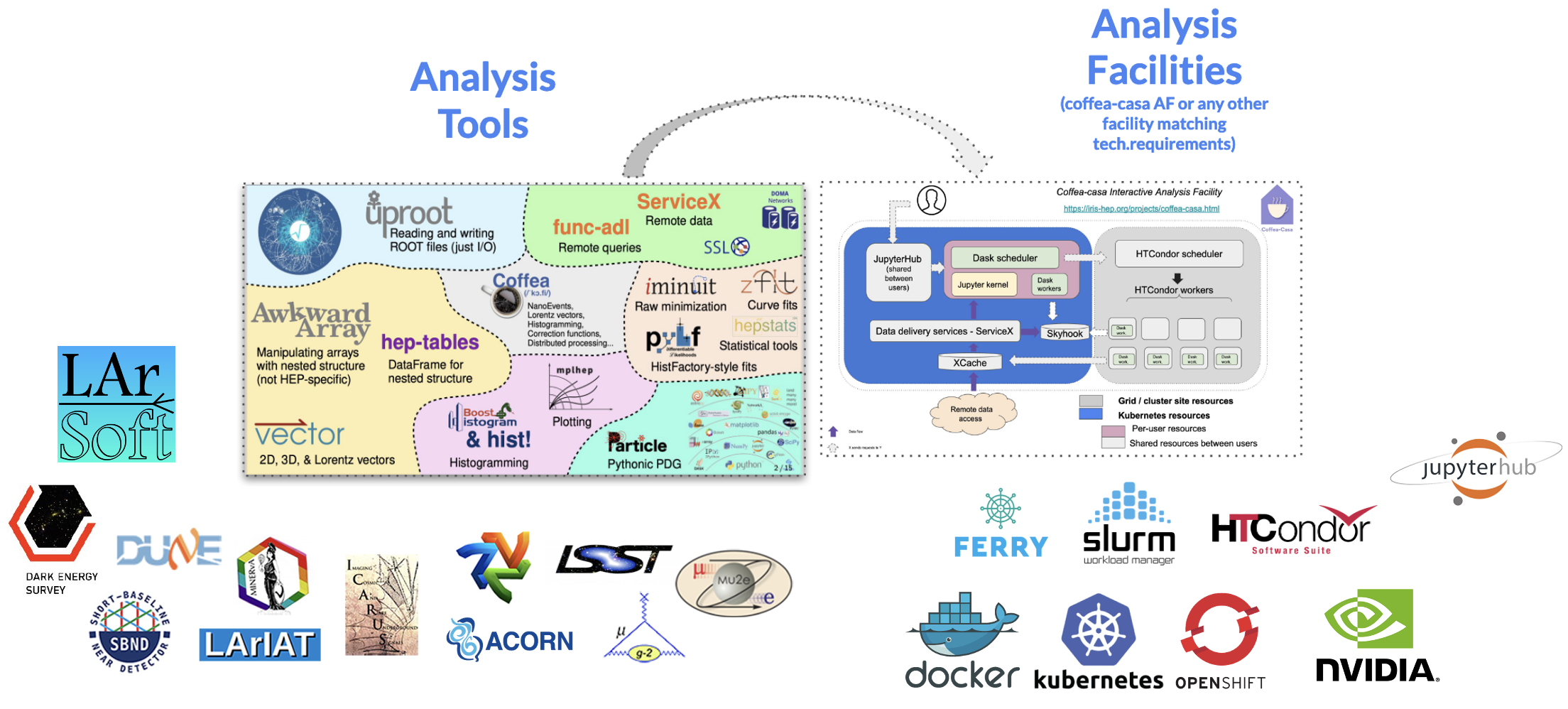}
    \caption{Diagram illustrating analysis facilities like Coffea-Casa and the EAF. In the left box are CMS relevant analysis tools included for the user and in the right box is an infrastructure diagram of the facility. Elements outside of the boxes illustrate some of the experiments using the facility as well as some of the back-end software included in the EAF.}
    \label{eaf-af-diagram}
\end{figure}

\subsection{The Elastic Analysis Facility (EAF)}
The EAF offers several unique features that distinguish it from more specialized analysis facilities, making it beneficial for future experimental collaborations. Unlike many other analysis facilities, the EAF is multi-VO, meaning it supports multiple experiments at Fermi National Accelerator Laboratory (Fermilab) including CMS, neutrino experiments, and accelerator development. As such, the EAF has the potential to serve as a unified research platform for the entire lab. Another distinction is the adherence to the security and authentication baselines required at a U.S. national laboratory. This partially drove the development of \textbf{HTCdaskgateway}, an extension of Dask Gateway, to connect EAF users to Fermilab's high-throughput batch systems while ensuring jobs are submitted under the user's identity, as required by the baseline. These features distinguish the EAF from university-style analysis facilities (such as Coffea-Casa) and other analysis platforms making it a possible template for multi-experiment institutions and colliders. Furthermore, it upholds the portability, reproducibility, usability, modernization, and sustainability of analysis facilities. 

In essence, the EAF functions as a comprehensive analysis ecosystem (Figure \ref{eaf-eco}), hosted on Kubernetes resources that are connected to grid and cluster resources at Fermilab. Users access experiment-specific analysis libraries and tools through a web-based JupyterHub interface, logging in via Fermilab single sign-on from the lab network or externally via proxy or VPN (Virtual Private Network).  The EAF offers multiple JupyterLab environments tailored to different experiments, including terminals, Jupyter notebooks, and machine learning software, as well as plug-ins like Dask, GitHub, and TensorBoard. Dask, a tool for executing distributed python code, is integrated with the Dask Gateway, allowing users to run and monitor their analyses on Fermilab computing resources via HTCondor\footnotemark[3]. 

\footnotetext[3]{HTCondor is a software system for high throughput computing that allows users to submit and execute jobs on remote machines within a defined computing environment configured by the software \cite{htcondor}.}

\begin{figure}[H]
    \centering
    \includegraphics[scale=0.25]{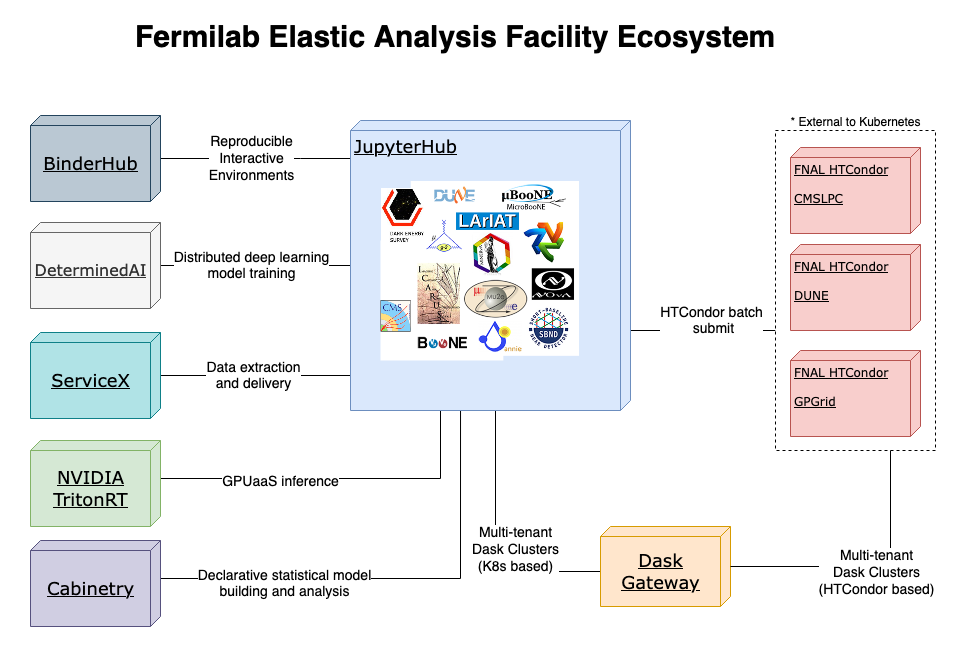}
    \caption{Diagram of the EAF ecosystem. At the center is JupyterHub,  which supports many Fermilab experiments, represented by the icons in the central box. JupyterHub is connected to services and software on the left while also interfacing with Fermilab HTCondor pools on the right via a Dask gateway.}
    \label{eaf-eco}
\end{figure}

\footnotetext[4]{OKD is a community driven Kubernetes distribution for application management \cite{okd-ref}.}

\footnotetext[6]{Docker is a platform that provides environments called containers that allow for packaging and running applications \cite{docker-ref}.}

\section{Back-End and Infrastructure}
The EAF is a complex ecosystem of applications and services. The JupyterHub instance is deployed as a \textbf{Kubernetes Deployment} on an OKD\footnotemark[4] cluster. The OKD cluster consists of a group of nodes with pooled resources, while a \textbf{Deployment} manages the life-cycle of assigned pods. A \textbf{pod} is the fundamental execution unit in Kubernetes, encapsulating one or more containers that share networking and storage resources. In the EAF, user notebooks are spawned as pods that run as the assigned user, orchestrated by the JupyterHub service. JupyterHub is configured using \textbf{Helm charts}, which are templatable YAML (Yet Another Mark-up Language) files which instantiate Kubernetes manifests on installation. The chart also specifies which container images are used for user notebooks. These images are Docker\footnotemark[6] images that have been created to meet experiment-specific requirements. There are foundational base images, which consist of the operating system and packages common to all notebooks. Each experiment has specific images which inherit from the base images, adding experiment-specific software and packages. Additionally, some experiments have variant derived images that inherit from these experiment-specific images. Each image has a specific tag propagated to the Helm chart, ensuring the correct images are used for the notebooks.

The OKD cluster consists of two types of nodes: an admin backplane, which manages the cluster, and worker nodes, which run JupyterHub, ancillary services, and user notebooks. The worker nodes are further subdivided into CPU nodes and GPU nodes, with the latter designated for executing GPU-intensive workloads. These nodes are interconnected via a 100 Gigabit network and are linked to a Ceph \cite{ceph-ref} storage cluster. The cluster includes 8 CPU worker nodes and 11 GPU worker nodes, which is sufficient for current operations, but more machines will be needed as usage scales to support more researchers. 

Table \ref{okd-cluster} below gives the node specifications.

\begin{table}[H]
    \centering
    \begin{tabular}{|c|c|c|c|c|}
    \hline
    Number of Nodes & Type of Node & Number of Cores & RAM (GB) & Network (Gb) \\ \hline \hline
    3 & CPU Worker & 78 & 354 & 100 \\ \hline
    2 & CPU Worker & 127 & 964 & 100 \\ \hline
    3 & CPU Worker & 126 & 472 & 100 \\ \hline
    3 & GPU Worker 4 A100 & 62 & 472 & 100 \\ \hline
    8 & GPU Worker 2 A100 & 127 & 477 & 100 \\ \hline
    \end{tabular}
    \caption{Specifications of the EAF worker and GPU nodes.}
    \label{okd-cluster}
\end{table}

It is important to note that these nodes are separate from those running Fermilab HTCondor jobs. The EAF has dedicated machines that run HTCondor jobs submitted from an EAF notebook. These dedicated machines are accessible to any experiment or single user using the EAF.

\section{Multi-Experiment Support}
Fermilab hosts a number of experiments and serves as the host lab for U.S.-CMS (US collaboration efforts on CMS). As such, there are several different types of workflows used with vastly different structures and tool requirements. The EAF was developed to support all of these analysis needs and currently supports CMS, the Deep Underground Neutrino Experiment (DUNE), the NuMI Off-Axis $\nu_{e}$ Appearance (NOvA) experiment, as well as various other neutrino experiments, astrophysics experiments, dark matter experiments, and accelerator development and monitoring efforts. To meet these needs, the EAF must provide researchers access to experiment-specific resources while ensuring usability for analysis. 

\subsection{Experiment Notebooks and Environments}
Each experiment has dedicated notebooks (both CPU and GPU as shown in Figure \ref{notebook-diagram}), based on the specialized container images described below. For clarity, a ``notebook'' here refers to the Jupyter instance running the specialized container. These container images follow the layered structure of base, experiment, and variant images. They are curated with input from both the research community and feedback from researchers. The EAF has 6 dedicated notebook groups and 1 generic notebook group, as shown in the diagram below.

\begin{figure}[H]
    \centering
    \includegraphics[scale=0.3]{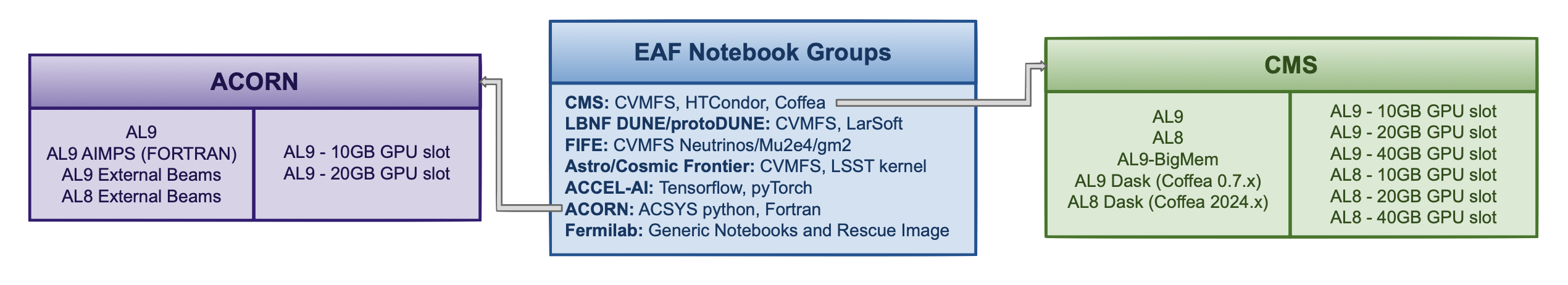}
    \caption{Diagram showing the EAF's experiment groups along with ACORN and CMS notebook offerings. Each experiment group is accompanied by a brief description of what is included in the notebooks. AL refers to Alma Linux, and the number indicates the version (e.g., AL9 refers to Alma Linux 9).}
    \label{notebook-diagram}
\end{figure}

Also, Figure \ref{notebook-diagram} shows the notebooks provided for CMS and ACORN (Accelerator Controls Operations Research Network), which highlight the differences in workflow needs. Many of the experimental groups have several variations of notebooks, as researchers in these groups have workflows that serve different purposes. The number of notebooks can be modified based on feedback from researchers. However, a notebook will not be added if it is intended for use only by a single researcher.

To enable different workflows, the notebooks are configured with experiment-relevant file systems, software, and tools. Each notebook has CVMFS (CernVM File System \cite{cvmfs-ref}) mounts that are specific to the experiment group. CVMFS is a read-only file system that gives the user access to data, storage and software. Because the mounts are experiment-specific, a CMS user does not have access to DUNE's CVMFS but will have access to CMS's CVMFS and vice-versa.  In addition, multiple programming languages are supported for experiment workflows. The EAF currently supports Fortran, C++ and Python. Other languages can also be supported if needed, but for interactive notebooks, the language must be compatible with JupyterHub (it must support a read-eval-print loop). This flexibility adds another layer of customization for experiment groups.

The notebooks also have commonly used packages and tools pre-installed for the experiment's researchers. Some experiments have these packages installed for all notebooks in the group, while others have some variations. For example, all DUNE notebooks install LArSoft, a software package for simulating liquid Argon time projection chambers as well as jobsub-lite (job submission software). On the other hand, only some CMS notebooks install Coffea (analysis framework software) and HTCdaskgateway (job submission software). The images with variations, like those for CMS, follow the same structure as the base/experiment/variant image system: a general experiment notebook with a baseline configuration (e.g. CVMFS), with variant notebooks adding additional tools such as Coffea. This is decided on an experiment-by-experiment basis, based on the needs communicated to the EAF. For all experiments, a GPU notebook is available to use that contains the same packages and configurations as the primary CPU notebook. Users can choose between 10GB, 20GB, and 40GB GPU slots. These notebooks are largely used for machine learning.

\subsection{Analysis on a Single Platform}
Another important feature of the EAF is its persistent shared storage, which ensures that users have access to all their files, no matter which notebook they choose. This allows a researcher to work on multiple projects and switch between them easily. Additionally, the EAF benefits from JupyterHub's features and extensions, which include easy file navigation, Git integration, user-friendly web interfaces, and convenient text editing. Figure \ref{eaf-screenshot} below provides a  snapshot of the EAF launcher, along with some of its interactive features. The EAF also supports terminals and Jupytext. Altogether, these features enable users to develop and perform analysis on a single platform, mitigating the need to configure environments, search for specific packages, or switch platforms for different projects. Instead, researchers can simply log on to the EAF, select an appropriate notebook, and begin working immediately. This approach is particularly beneficial for researchers and software developers working on multiple experiments, as it eliminates the need to change platforms.

\begin{figure}[H]
    \centering
    \includegraphics[scale=0.13]{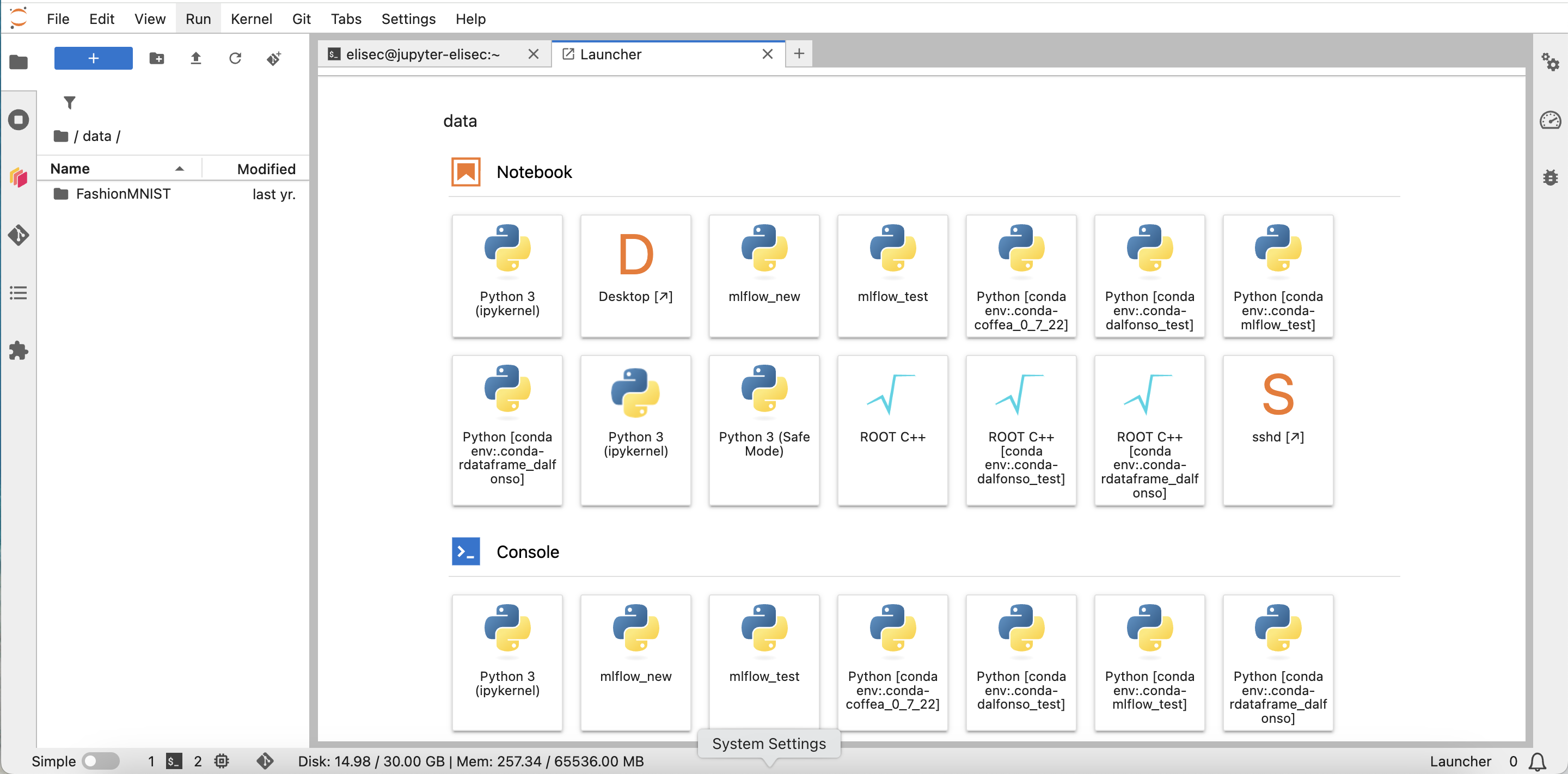}
    \caption{Snapshot of the EAF launcher. On the far left are the pre-installed extensions, including Git and Dask. Next to that is the file explorer. The launcher provides interactive Python notebooks, console, and (not shown here) other options such as Jupytext.}
    \label{eaf-screenshot}
\end{figure}

Analysis on a single platform also provides two key benefits for physics research: sustainability and portability. Even though it is quite complex, the EAF is a single platform to support and the configurations are standardized across each experiment. This means that the management and maintenance are centralized to Fermilab with responsibility for experiment environments delegated to researchers with software expertise and general feedback from researchers. This allows for a more sustainable approach to analysis that fosters collaboration. The second major benefit is portability. Researchers working on the same platform can more easily share their analyses for cross-checks or collaboration. A notebook runs the same container for all users, so a researcher can send a colleague their notebook, and it will run the same for both of them.

\section{Security and Authentication}
The EAF adheres to Fermilab's security guidelines, ensuring controlled access and encrypted connections. The lab's guidelines primarily require single sign-on (SSO) authentication, while the EAF itself remains behind the site firewall, accessible only via on-site networks, VPN, or proxy.

\subsection{Access at a US National Lab}
Fermilab employees, affiliates, and users are granted access to computing resources based on their roles and affiliations. Authentiation is managed through Fermilab's central SSO system, requiring a valid ``services'' account.  Off-site access is restricted to VPN (using services account credentials) or proxy (requiring both Kerberos and service account credentials). Additionally, the EAF enforces experiment-based access control. Users can access resources and notebooks associated with their experiment affiliation. For example, a CMS researcher is not permitted to use LBNF/protoDUNE. This access control is implemented via a central attribute repository called Ferry \cite{ferry-ref}. Ferry allows for services to query these attributes (e.g., experiment affiliation) and utilize them. In addition, the EAF deploys clients for creating VOMS (Virtual Organization Membership Service) proxies so that users can authenticate for their experiment's resources. Another key security feature is that EAF notebooks run under the assigned Unix UID (and supplemental GIDs) of the user. This allows seamless integration with external storage while maintaining standard Unix permissions, ensuring proper data access controls across different computing environments.

\subsection{HTCdaskgateway: Connecting HTCondor to Dask Gateway}
An important feature of analysis facilities is the ability to perform analysis through distributed computing. At Fermilab, this is primarily managed by HTCondor, but CMS is transitioning to incorporate Dask as part of their analysis framework, Coffea. Dask is widely used in ``big data'' applications and offers modern tools for scalable computing. While Dask natively supports HTCondor integration when users have access to the cluster backends, this is not the case for the EAF. Instead, users rely on Dask Gateway to connect to a Dask cluster \cite{Dask-gateway-site}. However, Dask Gateway does not support HTCondor in the same way. The default job submission mechanism in Dask Gateway is incompatible with Fermilab's batch systems: it attempts to submit jobs from the Dask scheduler, which runs as an ephemeral user, while the batch system requires that jobs must be submitted under the user's identity. This necessitated the development of a custom solution to connect CMS users to Dask and HTCondor through Dask Gateway via HTCdaskgateway, a Dask Gateway extension. 

HTCdaskgateway is a derived class of Dask Gateway that submits jobs on behalf of the user. It is available on pypi (Python package index) and is pre-installed on all CMS Dask notebook images, it simplifies the process of connecting Dask to HTCondor. When a user creates a new HTCdaskgateway instance in their python notebook, it configures the Dask scheduler, launches a Dask Gateway pod in Kubernetes, establishes necessary connections, and provides access to a zero-size Dask cluster.  The user can then scale the cluster as needed. Calling the \textit{scale} method in HTCdaskgateway generates an HTCondor Job Description Language (JDL) script and a start script. The JDL defines the job configuration, while the start script executes when the job starts on batch worker nodes, connecting the Dask workers to the scheduler within the proper container. Once scaled, the user now has their desired number of Dask-HTCondor workers ready to perform tasks from their notebook. 
Users can configure the worker image, requested worker memory, and requested worker cores. It is crucial that the worker image closely match the user's notebook (client) image, as Dask requires both its own version and the versions of its dependencies to be consistent between the worker and client. Additionally, the versions in the scheduler must match, but HTCdaskgateway ensures this automatically by using the same image for the worker and scheduler. The diagram in Figure \ref{HTCdaskgateway} below provides an overview of how HTCdaskgateway works.

\begin{figure}[H]
    \centering
    \includegraphics[scale=0.3]{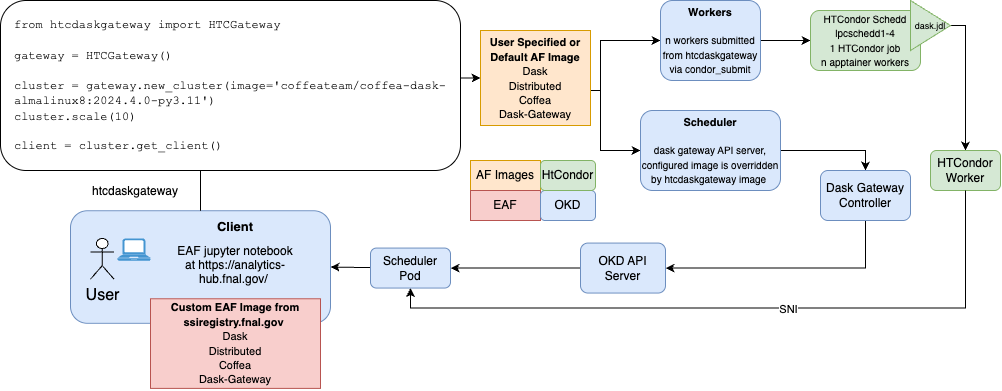}
    \caption{Diagram illustrating the HTCdaskgateway workflow. A user invokes HTCdaskgateway from the client to start workers and establish a connection with the scheduler. HTCdaskgateway submits the JDL from the client and connects to the HTCondor scheduler, this then spawns HTCondor workers that are connected to the Dask scheduler pod. The scheduler pod communicates with the client. At this point, the user sees Dask-HTCondor workers waiting for jobs.}
    \label{HTCdaskgateway}
\end{figure}

\subsubsection{What's Next}
There are two minor features in development for HTCdaskgateway. The first is the introduction of a dedicated check for experiment credentials. For a CMS user to submit jobs, valid credentials for experiment authentication are required. If they are not present, a generic error message prompts the user to check their proxy, which can be confusing for both developers and users. The new feature will implement a credentials check that will emit an error message tailored to the specific issue along with a suggested solution. The second feature focuses on the general improvement of error handling. Currently, errors messages returned to the user are vague and unhelpful. To address this, more precise error checking and clearer messages will be added to the gateway. Additionally, worker logs are not always provided to the user. A new mechanism will be added to ensure these logs, or at least their locations, are properly propagated back to the user. Other planned developments include an adaptive image default and additional worker configurations. The image default is hard-coded, but it would be more effective if it were dynamically set based on the client's image. 

A major planned development is the generalization of HTCdaskgateway. As it stands, the gateway is Fermilab-specific as well as CMS-specific. The CMS specificity is only in the default image, and the planned adaptive image default would address this. The Fermilab-specific nature stems from the manner in which the client connects to the Dask clusters, which will take more work to generalize. Once generalized, HTCdaskgateway could easily be used by other faciltiies that want to support similar levels of authentication and security requirements for Dask-HTCondor-Kubernetes implementations. Lastly, because the HTCdaskgateway is a derived class of Dask Gateway, it lacks some of the features and methods available in Dask Gateway itself. A subset of Dask Gateway features are implemented, but not all of them. For example, there is no way to auto-scale workers. This feature would need to be implemented in HTCdaskgateway for it to be available to users. This means that the gateway requires a lot more development and more user feedback to meet the same needs as Dask or Dask Gateway.

\section{EAF’s Advantages and Potential}
The EAF offers several advantages that will position it to handle analysis on exascale datasets in the near future. It modernizes the analysis process, bridges the software gap, and provides secure, standardized analysis templates. A central design principle of the EAF is the incorporation of current industry-standard software (when possible) to support the use, operations, and development of the facility. In terms of physics computing and software, the EAF represents an advance over older methods of analysis (e.g. using a remote machine via ssh and setting up environments to submit HTCondor jobs through scripts). By leveraging existing technologies like Dask and Kubernetes, the EAF brings the tools commonly used in ``big data'' into physics computing. These tools are more widely supported and benefit from an enhanced understanding of computer science and software development. As such, these newer tools are better able to make more efficient use of computational resources. The EAF offers great flexibility in software integration, which can evolve over time. This flexibility is enabled by the use of Kubernetes pods, containers, and images, which facilitate rapid updates and changes in a highly-available infrastructure. Kubernetes enables multiple experiment notebook with their own images. These images can be easily changed and updated to support various forms of analysis. For example, the CMS notebooks support both the ROOT \cite{root-ref} C++ analysis framework, and Coffea, a Pythonic analysis framework. The EAF has also demonstrated support for ROOT's parallel analysis framework RDataFrame via user-defined images and environments. Other benefits of modernizing include more sustainable software, more user friendly interfaces, better documentation, better security and authentication protocols, and more.

One key challenge that analysis facilities aim to address is the difficulty many physicists face in developing analyses. Physicists and physics students are not always well-versed in programming, software, or computing. Analysis facilities strive to bridge this gap, supporting researchers who may lack expertise to manage, develop, and prepare software and analyses. Developing an analysis traditionally takes a significant amount of time due to the amount of niche computing knowledge and domain-specific software still in use. The EAF provides a solution by offering  more intuitive user interfaces and an easier entry point with Python, a language  commonly known among physicists. Python is also more user-friendly than traditional C++, which requires a deeper understanding. Furthermore, the EAF is pre-configured so that beginning users can ramp up quickly. They can simply log in, open a notebook, customize it as needed, and start working. This allows computing professionals to manage the software integration and configuration with input from physicists. This results in a more usable and sustainable analysis platform. With the EAF, physicists no longer need specialized computing expertise to produce results.

However, the EAF is not without its challenges. The primary issues stem from user customizations, user support, and learning curves. User customizations can cause notebooks to crash or malfunction. There is also a clear need for more user support. Additionally, the EAF may also present a learning curve for those unfamiliar with Python-based interfaces. Each of these issues requires further consideration to find the most effective solutions for the EAF. Another significant challenge lies with the current mechanisms for establishing secure connections. The requirement for a Fermilab ``services'' account (as opposed to accepting a federated identity), along with the restrictions that the service is only accessible on-site or via VPN/proxy, creates additional barriers to entry. These access limitations can make it difficult for users who are off-site or lack the necessary credentials to engage with the EAF.  Improving these access methods will be key to making the EAF accessible to a broader user base.

\section{The EAF and The Future}
The EAF is in active use, and this usage is expected to grow with upgrades to existing experiments and as new experiments are introduced. With secure multi-experiment support, it offers significant benefits to any future collider or multi-experiment institution, offering a scalable and adaptable framework for their needs. Preparation for HL-LHC also enhances the EAF's potential for the future, as it aims to facilitate fast analysis of large data volumes. As the EAF progresses to this goal, it will increasingly support the analysis objectives of future colliders. 

\subsection{EAF's Potential for Future Colliders}
A significant benefit of an EAF-style facility is the ability to support multiple experiments. As demonstrated at Fermilab, the EAF is able to effectively meet the needs of a diverse set of experiments, allowing any researcher to utilize EAF resources and produce results. For future colliders, a unified platform capable of supporting all beamline experiments would be particularly valuable. A centralized analysis facility would streamline documentation, operations, and maintenance, reducing the need for each experiment to develop and sustain its own software and computing infrastructure. In addition, the platform would support beamline operations and accelerator development, which benefits all experiments. As at Fermilab, researchers at a future collider could rely on a single, well-maintained system for conducting their analyses. Also, a shared facility fosters better collaboration, as experiments work together to maintain and improve their parts of the facility. Another promising feature of the EAF can provide is the HTCdaskgateway. While still under development, this tool could be adapted for a variety of job submission frameworks besides HTCondor, including future batch submission systems. Regardless of the underlying processing system, a future collider would be able to use the gateway to connect Kubernetes resources to high-throughput computing resources, maintaining flexibility as technologies evolve.

Furthermore, as discussed in ``EAF’s Advantages and Potential'', many of the EAF's advantages persist over time due to its flexibility. The use of Kubernetes allows software to be easily updated or replaced. With dedicated development teams, modernizing can be a continuous process, enabling the integration of new tools and services relevant to future colliders. This adaptability is a core principle for the EAF. 
As computing and physics continue to diverge into increasingly specialized fields, the software gap will also continue to widen. Future colliders will benefit greatly from a platform designed to be intuitive, reducing the time users spend navigating complex or unfamiliar computing tools. By relying on dedicated teams of software engineers and computing specialists, the platform can be continuously managed, modified, and improved by experts in the field.  Another advantage of this approach is long-term sustainability. A centralized analysis facility, with well-documented infrastructure and input from physicists, helps mitigate knowledge loss when key personnel retire. It also prevents the platform from becoming obsolete. As long as modernization efforts continue and the facility is managed by computing professionals, this platform can remain a reliable and effective resource for future colliders.

\subsection{HL-LHC and the EAF}
In the near future, the EAF is expected to play a significant role in the next generation of analysis for the HL-LHC. Several aspects of the EAF were specifically designed with HL-LHC requirements in mind to support modern analysis techniques for large data volumes. To meet those demands, the EAF is committed to developing a high-performance, high-throughput facility by continuously exploring and implementing modern computing technologies. As previously highlighted, modernization brings benefits in both usability and performance, making the EAF increasingly beneficial to other experiments requiring large-scale data processing.  

Progress toward HL-LHC analysis within the EAF is tracked through the IRIS-HEP (Institute for Research and Innovation in Software for High Energy Physics) Analysis Grand Challenge (AGC). The AGC is an initiative aimed at facilitating and documenting next-generation analysis methods for HL-LHC, ensuring that analysis systems not only meet the data processing requirements of the HL-LHC but also offer more advanced capabilities than current systems \cite{irishep-site}. The EAF has actively participated in AGC events and has been tested using a standardized notebook. Following recent AGC tests with Coffea-Casa, a dedicated throughput test is being planned to assess the EAF's current capabilities and identify potential areas for improvement. Additionally, benchmarking studies have been conducted to evaluate EAF performance using the standard AGC analysis. Some of these results are shown in Figures \ref{benchmarks} below. Currently, there are no established benchmarks defining what constitutes an optimal analysis facility. However, analyses like these will help shape AGC expectations and establish meaningful performance comparisons across analysis facilities. As development progresses, the EAF remains on track to provide a high-performance analysis platform for the HL-LHC and potentially for future institutions and colliders. 

\begin{figure}[H]
    \centering
    \begin{subfigure}{0.45\textwidth}
        \centering\captionsetup{width=.95\linewidth}%
        \includegraphics[width=0.95\linewidth]{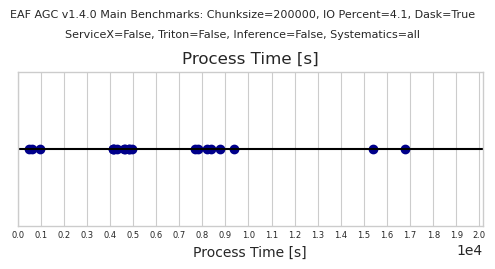}
        \caption{Processing time (in seconds) for the v1.4.0 \cite{agc-v1.4.0} AGC notebook. The processing time represents the duration from from start to finish of data processing as recorded by Dask.}
    \end{subfigure}%
    \hfill
    \begin{subfigure}{0.45\textwidth}
        \centering\captionsetup{width=.7\linewidth}%
        \includegraphics[width=0.95\linewidth]{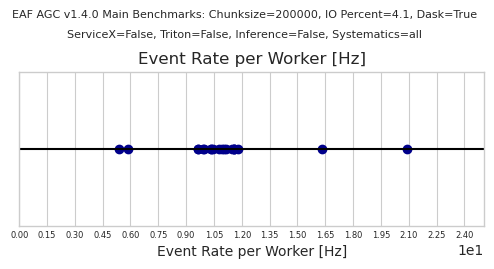}
        \caption{Event rate per worker (in Hertz), calculated by dividing the number of events by the processing time.}
    \end{subfigure}
\end{figure}

\begin{figure}[H]
\ContinuedFloat
    
\begin{subfigure}{0.45\textwidth}
    \centering\captionsetup{width=.8\linewidth}%
    \includegraphics[width=0.95\linewidth]{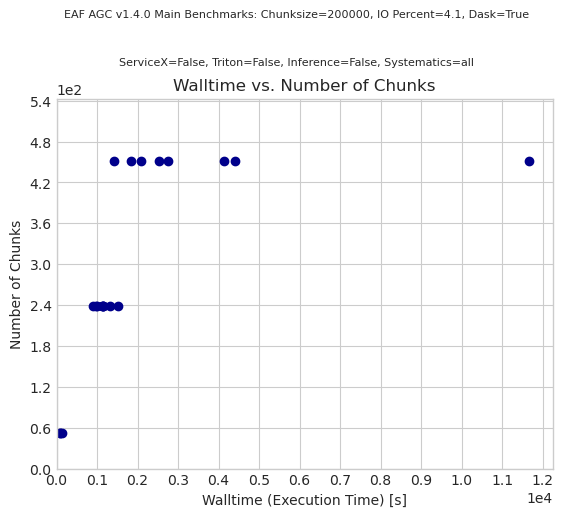}
    \caption{Walltime (in seconds) vs. number of data chunks. Walltime refers to the total time taken by workers from start to finish for data processing, measured using Python timing, while chunks represent discrete units of input data.}
\end{subfigure}% 
\hfill
\begin{subfigure}{0.45\textwidth}
    \centering\captionsetup{width=.95\linewidth}%
    \includegraphics[width=0.95\linewidth]{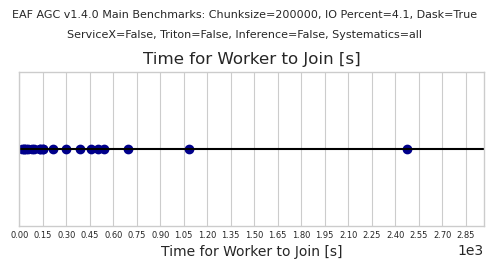}
    \caption{Join time for workers on the EAF. Join time refers to the latency between requesting workers through the HTCdaskgateway and their availability for performing work. The primary component is the delay in the batch system between job submission and start.}
\end{subfigure}
        
\caption{Examples of benchmarks for the EAF.}
\label{benchmarks}
\end{figure}

\section{Conclusions}
The Elastic Analysis Facility (EAF) is a powerful and versatile analysis platform with significant potential for future colliders and their evolving needs. It securely supports multiple experiments, integrating experiment-specific tools and user-friendly interfaces. By leveraging the expertise of software and computing professionals, the EAF helps bridge the software gap while enhancing performance and usability through a modernized analysis suite. Designed with the HL-LHC in mind, the EAF is built for high-throughput processing to handle large data volumes efficiently. Overall, the key advantages of a facility like the EAF are portability, reproducibility, usability, and sustainability. These qualities make it a potential template for other labs, research facilities, future colliders, and multi-experiment institutions seeking a robust, scalable analysis platform for both current and future generation of scientists.

\section{Acknowledgments}
This work was supported by the Department of Energy grants DE-SC0023528 and DE-SC0017647. This document was prepared by CMS using the resources of the Fermi National Accelerator Laboratory (Fermilab), a U.S. Department of Energy, Office of Science, Office of High Energy Physics HEP User Facility. Fermilab is managed by FermiForward Discovery Group, LLC, acting under Contract No. 89243024CSC000002.

\end{document}